\documentclass{comjnl}

\usepackage{tabulary,graphicx,times,caption,fancyhdr,amsbsy,latexsym}
\usepackage[utf8]{inputenc}
\usepackage{url,multirow,morefloats,floatflt,cancel,tfrupee,textcomp,colortbl,xcolor,pifont}
\usepackage{soul}
\usepackage{amsmath,amssymb,amsfonts,amsthm,bm}
\usepackage[retainorgcmds]{IEEEtrantools}
\usepackage{booktabs}
\usepackage{algorithm}
\usepackage{algorithmic}
\usepackage{bigstrut}
\usepackage{diagbox,color}
 
\usepackage{bbding}

\DeclareMathOperator*{\Proj}{Proj}
\DeclareMathOperator*{\sign}{sign}
\DeclareMathOperator*{\softmax}{softmax}
\DeclareMathOperator*{\Dist}{Dist}
\DeclareMathOperator*{\argmin}{arg min}
\DeclareMathOperator*{\argmax}{arg max}

\let\oldmaketitle\maketitle
\renewcommand{\maketitle}{\oldmaketitle\setcounter{footnote}{0}}

\begin{document}

\title[Evaluating Membership Inference Through Adversarial Robustness]{Evaluating Membership Inference Through Adversarial Robustness}

\author{Zhaoxi Zhang} \affiliation{School of Computer and Information Science, Southwest University}
\author{Leo Yu Zhang} \affiliation{School of Information Technology, Deakin University}
\author{Xufei Zheng \Envelope$^,$}\email{zxufei@swu.edu.cn} \affiliation{School of Computer and Information Science, Southwest University}
\author{Bilal Hussain Abbasi} \affiliation{School of Information Technology, Deakin University}
\author{Shengshan Hu} \affiliation{School of Cyber Science and Engineering, Huazhong University of Science and Technology}

\shortauthors{
    Z. Zhang \textit{et al.}
}

\received{00 January 2022}
\revised{00 Month 2022}

\keywords{
    deep learning; privacy leakage; membership inference attack; adversarial attack
}

\begin{abstract}
The usage of deep learning is being escalated in many applications. 
{Due to its outstanding performance, it is being used in a variety of security and privacy-sensitive areas in addition to conventional applications.}
One of the key aspects of deep learning efficacy is to have abundant data. This trait leads to the usage of data which can be highly sensitive and private, which in turn causes wariness with regard to deep learning in the general public.
Membership inference attacks are considered lethal as they can be used to figure out whether a piece of data belongs to the training dataset or not. This can be problematic with regards to leakage of training data information and its characteristics. 
To highlight the significance of these types of attacks, we propose an enhanced methodology for membership inference attacks based on adversarial robustness, by adjusting the directions of adversarial perturbations through label smoothing under a white-box setting. 
We evaluate our proposed method on three datasets: Fashion-MNIST, CIFAR-10, and CIFAR-100. Our experimental results reveal that the performance of our method surpasses that of the existing adversarial robustness-based method when attacking normally trained models. 
Additionally, through comparing our technique with the state-of-the-art metric-based membership inference methods, our proposed method also shows better performance when attacking adversarially trained models. 
The code for reproducing the results of this work is available at \url{https://github.com/plll4zzx/Evaluating-Membership-Inference-Through-Adversarial-Robustness}. 
\end{abstract}
\maketitle

\section{Introduction}
\label{Sec:Intro}
Deep learning is being used in many fields \cite{Chan2016ListenAA,Lawhern2018EEGNetAC,Karras2019ASG,subbarayalu2019hybrid}, some of these areas are security and privacy sensitive, such as face recognition \cite{KemelmacherShlizerman2016TheMB, Schroff2015FaceNetAU}, medical diagnosis \cite{Burlina2017AutomatedGO, Kourou2015MachineLA},
code analysis \cite{qiu2021survey,lin2020software,chen2019android}, security incident prediction \cite{sun2019data,liu2018detecting}, and intrusion detection \cite{subbarayalu2019hybrid,gogoi2014mlh}.
In the deep learning domain, a large amount of data generally leads to better performance. Therefore, it is common to use a huge amount of data to train deep learning models. Although this trait of deep learning can result in better performance, there can also be some severe consequences. For instance, the usage of data that contains private and sensitive information can ultimately lead to privacy issues. Such privacy risks associated with deep learning start to raise ethical and security concerns among the society \cite{EUdataregulations2016, hipaa}.

One of the most studied attacks for privacy leakage is known as the membership inference attack. 
The adversary's goal in this type of attack is to figure out if a certain piece of data was utilized in training or not. 
For example, if a model has been trained using patient data, one can use the membership inference attack to associate a particular patient with diseases of certainty if the attack reveals that the very piece of patient data has been used for training.

Various studies show that deep learning models are vulnerable to such privacy attacks \cite{Shokri2017MembershipIA,Yeom2018PrivacyRI,Nasr2019ComprehensivePA,Leino2020StolenML,Sablayrolles2019WhiteboxVB, Song2020SystematicEO,Li2020MembershipLL,ChoquetteChoo2020LabelOnlyMI}. The work in \cite{Shokri2017MembershipIA} first suggested using a neural network to perform membership inference as a binary classification task via the black-box access of the victim neural model. It is further demonstrated in \cite{Yeom2018PrivacyRI} that membership inference is mainly caused by overfitting, and some metrics and thresholds were proposed for better inference. Subsequent works \cite{Nasr2019ComprehensivePA,Leino2020StolenML,Sablayrolles2019WhiteboxVB,Song2020SystematicEO} further evaluated privacy risks caused by membership inference in a white-box manner. Furthermore, the work in \cite{Song2020SystematicEO} revealed that by using suitable metrics, metric-based attacks can achieve competitive performance for membership inference when compared with neural classifier \cite{Shokri2017MembershipIA}. In this paper, we focus on metric-based white-box membership inference attacks for classification neural models.

From another line of research, adversarial examples of deep neural models have been extensively studied \cite{Szegedy2014IntriguingPO,MoosaviDezfooli2016DeepFoolAS,Li2020MembershipLL,ChoquetteChoo2020LabelOnlyMI,zhang2021self,hu2021advhash}. The seminal work in  \cite{Szegedy2014IntriguingPO} made use of elaborately crafted adversarial perturbations, which are very tiny and unnoticeable to human eyes, to cause misclassifications of a victim model.
It was discovered in \cite{MoosaviDezfooli2016DeepFoolAS} that adversarial perturbations can be determined as the distance between data points and decision boundaries. 
The works \cite{Li2020MembershipLL,ChoquetteChoo2020LabelOnlyMI} bridged membership inference attacks and adversarial attacks by suggesting adversarial robustness as a metric for membership inference.

These methods are based on the fact that a larger perturbation is needed to construct an adversarial example from a member than from a non-member data. As pointed out by \cite{MoosaviDezfooli2016DeepFoolAS}, the magnitude of the adversarial perturbation can be considered as a metric to measure adversarial robustness, (i.e., the member is more robust than the non-member in terms of adversarial perturbation).

Associating this fact with inference attack, the magnitude of the adversarial perturbation can be also used as the metric in membership inference, i.e., the larger the perturbation an example needs in an adversarial attack, the more likely it is a member in an inference attack. 
In contrast, traditional metric-based membership inference attacks only use the information collected from the output of the victim model to infer whether a piece of data is a member or not. 
But adversarial robustness-based attacks are not restricted to the output of the victim model. In particular, adversarial robustness-based attacks can collect information from the victim model through multiple iterations while generating adversarial examples. 
This makes adversarial robustness-based methods more convenient and powerful than metric-based methods.

This work moves one step further towards the study of membership inference attacks by making use of the notion of adversarial robustness. In the case of targeted adversarial perturbation generation, target labels represent directions of perturbation. 
Therefore, the goal of attackers in adversarial attacks is to find the smallest perturbation that makes the example cross the decision boundary in a specified direction. 
This direction is commonly adopted as the one-hot encoding of the target label. 
On the contrary, in adversarial robustness-based membership inference attacks, the main objective is to find perturbations that can distinguish members and non-members, which is independent of the task of finding the smallest perturbation. 
In other words, the smallest perturbation size is not a concern in membership inference attacks, but it is a key factor for the success of adversarial attacks.

\begin{figure}[t]
    \centering
    \includegraphics[width=6cm]{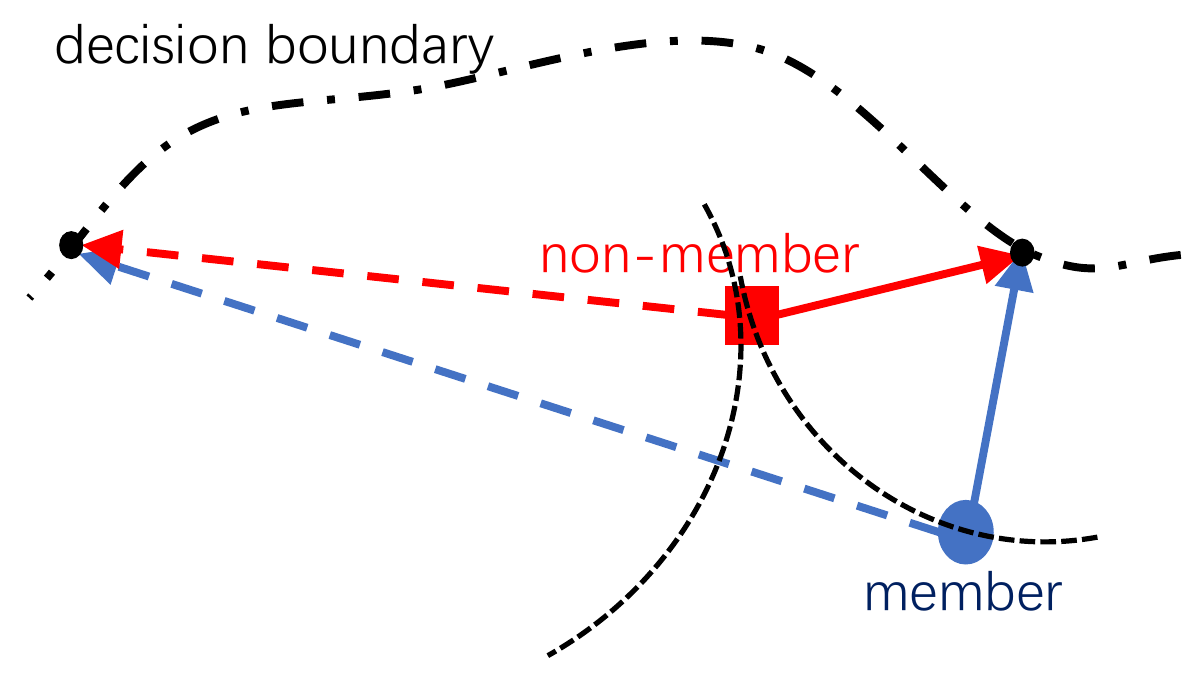}
    \caption{
        An illustration of directional distance. In the left direction, to cross the decision boundary, the size of adversarial perturbations needed for the non-member and member is different. In the right direction, the size of adversarial perturbations is the same. 
    }
    \label{Fig:direction}
\end{figure}

Due to this fact, the gap between members and non-members could be different if different adversarial directions (other than one-hot encoding) are used, as illustrated in Fig~\ref{Fig:direction}. 
Based on this observation, we propose a membership inference attack by enlarging the gap between the adversarial directional distance of members and non-members.
With the proposed method, we discuss the upper bound of membership inference by comparing it with the state-of-the-art inference attacks. 

The contributions of this paper are two-fold:
\begin{itemize}
    \item We propose a new technique to enhance the existing adversarial robustness-based membership inference methods by adjusting the directions of adversarial perturbations through label smoothing.
    \item We evaluate the performance of our proposed method on normally/adversarially trained models and compare it with metric-based methods. Experimental results demonstrate that the proposed method outperforms metric-based methods, which suggests that our technique is more lethal, i.e., it can result in stronger inference attack. 
\end{itemize}

The rest of the paper is organized as follows. The background knowledge is discussed in Section~\ref{sec:background}. 
Section~\ref{sec:mbmi} presents the details of classic metric-based membership inference attacks that will be used for comparison purpose. 
Section~\ref{sec:adv_mi} presents the details of the proposed adversarial robustness based membership inference and the directional distance based membership inference methods. 
Section~\ref{sec:exp} presents comprehensive experiments on the methods shown in Secs.~\ref{sec:mbmi} and \ref{sec:adv_mi}, and the conclusion is drawn in Sec.~\ref{sec:conclusion}.

\section{Background}
\label{sec:background}

\subsection{Deep Learning Basics}

Similar to literature studies on membership inference \cite{Nasr2019ComprehensivePA,Leino2020StolenML,Sablayrolles2019WhiteboxVB,Song2020SystematicEO}, this paper also focuses on deep neural network models used for (image) classification. 
For a trained neural model $F$, it classifies its input example $x$ by 
\begin{IEEEeqnarray}{rCl}
    y &=& F(x)  = \softmax(Z(x)),
\end{IEEEeqnarray}
where $Z$ is the logits, and $y$ is a probability vector with its largest entry (i.e., $\max (y)$) called the  confidence score.
For the training of $F$, two kinds of labels, one-hot label and smoothed label, can be used. If an example belong to class $k$ ($k\in [1,n]$), one-hot encoded label is characterized by 
\begin{IEEEeqnarray}{C}
    p_n(i): (0, \cdots, 0, 1, 0, \cdots, 0), \nonumber \\
    k=\argmax_{i\in[1,n]}(p_n(i)).
    \label{eq:one_hot}
\end{IEEEeqnarray}
Smoothed label (i.e., label smoothing technique) was firstly introduced by \cite{Szegedy2016RethinkingTI} to improve model performance and {was later discovered to be very useful for information distillation \cite{Hinton2015DistillingTK}}. It flattens the confidence of one-hot label by using
\begin{IEEEeqnarray}{C}
    q_n(\lambda, i): (\frac{1-\lambda}{n-1}, \cdots, \frac{1-\lambda}{n-1}, \lambda, \frac{1-\lambda}{n-1}, \cdots, \frac{1-\lambda}{n-1}), \nonumber \\
    k=\argmax_{i\in[1,n]}(q_n(\lambda, i)),
    \label{eq:label_smoothing}
\end{IEEEeqnarray}
where $\lambda$ is the new confidence score of the smoothed label $q_n(\lambda, i)$ if $\lambda \in (\frac{1}{n},1]$.

\subsection{Adversarial Examples and Adversarial Training}
\label{subsec:aeat}

In the case of classification tasks, adversarial examples are crafted with the purpose of misleading classification after a model is well-trained and deployed.
Adversarial examples can be easily generated by adding small perturbations to original data, which can be formulated as follows:
\begin{IEEEeqnarray}{rCl}
                       && \hat{x}_i = x_i+ \delta_i, \| \delta_i \|_{p} < \epsilon,\\
    \mathrm{untargeted~attack:~} && \argmax F(x_i) \neq \argmax F(\hat{x}_i), \IEEEeqnarraynumspace \\
    \mathrm{targeted~attack:~}   && F(\hat{x}_i) = y_t \mathrm{~and~}  F(x_i) \neq y_t ,
\end{IEEEeqnarray}
where $\hat{x}_i$ represents the adversarial version of the benign example $x_i$, $\delta_i$ is the adversarial perturbation,  $\epsilon$ is the upper bound of the allowable perturbation and $y_t$ is the targeted label. Hereinafter, we take the targeted version to illustrate different adversarial attacks, though all adversarial attacks can be also used for untargeted purpose.

FGSM \cite{Goodfellow2015ExplainingAH} is the first and most widely used method for generating adversarial examples as it is simple but very efficient. Adversarial examples generated by FGSM can be formulated as follows:
\begin{equation}
    \hat{x}= x+\epsilon \cdot \sign(\nabla_x \mathcal{L}_{CE}(x,y_t))).
\end{equation}
\noindent
PGD \cite{Madry2018TowardsDL} is proposed to improve FGSM through iterating the core component of FGSM. During each iteration, it is ensured that the generated example does not exceed the norm restriction by making use of a projection operator $\Proj$. The iteration formula of PGD is as follows: 
\begin{equation}
    \hat{x}^m= \Proj (\hat{x}^{m-1}+\epsilon \cdot \sign(\nabla_x \mathcal{L}_{CE}(x,y_t)))
\end{equation}
\noindent
C\&W \cite{Carlini2017TowardsET} is the sate-of-the-art adversarial example generation method. C\&W attempts to get the optimal perturbation in a iterative manner. C\&W is formulated as 
\begin{IEEEeqnarray}{c}
    \argmin_{\hat{x}_i} \ \alpha \cdot \mathcal{L}_{CE}(F(\hat{x}_i), y_t) + \|\hat{x}_i-x_i\|_p, 
\end{IEEEeqnarray}
where $\alpha$ is a hyperparameter used to adjust the trade-off between two loss functions when generating adversarial examples. Since C\&W directly optimizes $\|\hat{x}_i-x_i\|_p$, adversarial perturbations generated by this method are usually smaller than methods like FGSM and PGD.

PGD adversarial training (PGD-AT) \cite{Madry2018TowardsDL} is one of the most widely used method for defending against adversarial examples. Basic idea of this method is to first produce adversarial examples generated by PGD attack, and then rectify the adversarial examples to its original benign label $y$ to augment the training dataset. PGD-AT can be  formulated as the ($\min\max$) problem below:
\begin{equation}
    \min_{\theta} \max_{\hat{x} \in \mathcal{B}_{\epsilon}(x)} \mathcal{L}(F_{\theta}(\hat{x}), y),
\end{equation}
where $\mathcal{B}_{\epsilon}(x)$ contains all the adversarial examples corresponds to $x$ bounded by $\epsilon$. By doing so, the resultant trained model owns enhanced robustness against adversarial attacks.

\subsection{Privacy Leakage and Membership Inference}
\label{subsec:plmi}
As mentioned above, membership inference attack aims to figure out whether a piece of data belongs to the training dataset or not. 
There are two types of attack settings: black-box and white-box. 
In the black-box setting, attackers can only access the output of target model. In this type of attack, the most famous technique is to first train a shadow model, which essentially duplicates the functionality of the target model, and then perform inference attack on the shadow model \cite{Shokri2017MembershipIA}. 
In contrast, in the white-box setting, attackers can also access internal details of the target models \cite{Nasr2019ComprehensivePA, Leino2020StolenML, Sablayrolles2019WhiteboxVB, Song2020SystematicEO}. 
Typically, the white-box attack is stronger than the black-box. 
This is due to the fact that adversary has access to model parameters and neuron activations of the model in such attacks. 
However, black-box attacks can also perform well given that the attack is designed carefully and systematically. For instance, the work \cite{Song2020SystematicEO} showed the performance of black-box attacks is close to white-box attacks under some attack settings.

Recent research suggests that membership inference attacks can be linked with the well-known phenomenon in deep learning: model overfitting \cite{Yeom2018PrivacyRI}. The rationale behind this is overfitted model
can lead to significant differences between members and non-members under a variety of measurements. 
For example, overfitted model prefer higher confidence score, lower entropy, and smaller values of loss function for member examples. It is easy for the attacker to make use of such significant differences to differentiate members and non-member, which lead to the popularity of metric-based inference attacks \cite{Shokri2017MembershipIA,Yeom2018PrivacyRI,Song2020SystematicEO}.

\section{Metric-Based Membership Inference}
\label{sec:mbmi}

Following the discussions in Sec.~\ref{subsec:plmi}, we present the details of renowned metric-based membership inference methods here. 
Metric-based membership inference use certain designed metrics to measure the differences between members and non-members and these attacks heavily rely on model overfitting \cite{Yeom2018PrivacyRI}. 
{Common representative metric-based attacks \cite{Nasr2019ComprehensivePA,Leino2020StolenML,Sablayrolles2019WhiteboxVB, Song2020SystematicEO}} are summarized as follows. 

\noindent 
\textbf{The output of model is correct or not}: This inference method simply take examples with correct model outputs (i.e., correct classification) as members and examples with incorrect outputs as non-members. Due to its simplicity, it is commonly used as the baseline for evaluating inference attack and it can be formulated as 
\begin{IEEEeqnarray}{C}
    I_{bl} (x) = \mathbb{I}\Big( \argmax F(x)= k \Big), 
\end{IEEEeqnarray}
where $k \in [1,n]$ is the true class $x$ belongs to, and $\mathbb{I}()$ represents the indicator function that equals to one when the statement inside is true\footnote{The example $x$ is inferred as a member when $\mathbb{I}(\cdot)=1$.}  and zero when is false.

\noindent 
\textbf{The confidence score of output}: This inference method uses the confidence score as the metric for inference. The underlying rationale is examples in the training set (members) normally have higher confidence scores. The method $I_{cfs}$ is defined as 
\begin{IEEEeqnarray}{C}
    I_{cfs} (x) = \mathbb{I}\Big( \max F(x)>\tau \Big),
\end{IEEEeqnarray}
where $\tau$ is a predefined threshold\footnote{We abuse the notion $\tau$ as the threshold for different metric-based methods hereinafter, as it will not cause any ambiguity. }.

\noindent 
\textbf{The cross entropy of output}: Similarly to the methods above, the examples in training set normally has lower cross entropy loss since the cross entropy associated with members will be minimized during model training. The method $I_{CE}$ is defined as 
\begin{IEEEeqnarray}{C}
    I_{CE} (x) = \mathbb{I}\Big( CE(F(x), y)<\tau \Big).
\end{IEEEeqnarray}

\noindent 
\textbf{The entropy and m\_entropy of output} \cite{Song2020SystematicEO}: After training, the model output $F(x)$ of a member example $x$ will be optimized to match its one-hot encoded label $y$, which reduces the  
entropy contained in $F(x)$. So, $I_{entropy}$ is defined as
\begin{IEEEeqnarray}{C}
    I_{entropy} (x) = \mathbb{I}\Big( \mathrm{entropy}(F(x))<\tau \Big).
\end{IEEEeqnarray}
To make the metric monotonic with respect to the confidence score (i.e., prediction probability), the work in \cite{Song2020SystematicEO} defined $\mathrm{m\_entropy}$ as
\begin{IEEEeqnarray}{rCl}
    \mathrm{m\_entropy}(x, k) &=& -(1-F(x)_k) \log(F(x)_k) \nonumber \\
                              & & -\sum_{i \neq k}F(x)_i \log(1-F(x)_i), 
\end{IEEEeqnarray}
where $k \in [1,n]$ is the true class $x$ belongs to. By making use of $\mathrm{m\_entropy}$, the inference method $I_{m\_entropy}$ is defined as 
\begin{IEEEeqnarray}{rCl}
    I_{m\_entropy} (x, y) &=& \mathbb{I}\Big( \mathrm{m\_entropy}(F(x), y)<\tau \Big). 
\end{IEEEeqnarray}

\section{Adversarial Robustness Based Membership Inference}
\label{sec:adv_mi}
This section presents the details of how to use adversarial robustness to build new metric-based membership inference attacks. 
As discussed in Sec.~\ref{Sec:Intro}, the key observation is that member examples tend to be more robust (i.e., larger adversarial perturbation) than non-members when constructing their adversarial counterparts \cite{Li2020MembershipLL, ChoquetteChoo2020LabelOnlyMI}. 
By probing in different adversarial directions, it is possible to separate adversarial perturbations associated with member and non-member data better than those from the default direction.
We proceed with our discussion by first introducing a strawman inference approach that straightforwardly employs this observation, and then figuring out how to find the optimal directional adversarial perturbation to build stronger inferences. 

\subsection{A Strawman Inference Approach}
Due to a lot of reasons like the randomness contained in the optimization process, the semantic of the selected benign example and etc., an adversarial attack may fail to produce a lethal adversarial example. That said, after running a given adversarial attack, for example, PGD, on a benign example $x$, the resultant $\hat{x}$ still belongs to the same class as $x$. 

Considering members are generally more robust in adversarial attacks, we suggest the following strawman inference approach: 
\begin{IEEEeqnarray}{rCl}
    I_{sm\_adv} (x) = \mathbb{I}\Big( \argmax F(x)= \argmax F(\hat{x}) \Big) .
\end{IEEEeqnarray}
In this approach, if the adversarial attack fails on a given example $x$, $I_{sm\_adv}$ takes $x$ as a member, and vice versa.

\begin{table*}
    \centering
    \begin{minipage}[b]{.45\textwidth}
        \centering
        \caption{
            Average distance for normally trained model on Fashion's member dataset.
        }
        \resizebox{0.8\textwidth}{!}{
        \begin{tabular}{r|rrrr}
        \toprule
        \diagbox[width=12mm, height=6mm]{$T$}{$\lambda$} & 0.4   & 0.6   & 0.8   & 1 \\
        \midrule
        0.5   & 2.538 & 2.481 & 2.475 & 0.854 \\
        1     & 1.600   & 1.550  & 1.542 & \underline{0.653} \\
        3     & 0.856 & 0.889 & 0.887 & 0.592 \\
        5     & 0.705 & 0.756 & 0.758 & 0.664 \\
        \bottomrule
        \end{tabular}%
        }
        \label{tab:fashion_ptb_normal_train}%
    \end{minipage} \quad \quad
    \centering
    \begin{minipage}[b]{.45\textwidth}
        \centering
        \caption{
            Average distance for normally trained model on Fashion's non-member dataset.
        }
        \resizebox{0.8\textwidth}{!}{
        \begin{tabular}{r|rrrr}
        \toprule
        \diagbox[width=12mm, height=6mm]{$T$}{$\lambda$} & 0.4   & 0.6   & 0.8   & 1 \\
        \midrule
        0.5   & 2.546 & 2.495 & 2.479 & 0.849 \\
        1     & 1.606 & 1.573 & 1.557 & \underline{0.653} \\
        3     & 0.850  & 0.878 & 0.877 & 0.591 \\
        5     & 0.711 & 0.759 & 0.758 & 0.660 \\
        \bottomrule
        \end{tabular}%
        }
        \label{tab:fashion_ptb_normal_test}%
    \end{minipage}
    \centering
    \begin{minipage}[b]{.45\textwidth}
        \centering
        \caption{
            Average distance for normally trained model on CIFAR-10's member dataset.
        }
        \resizebox{.8\textwidth}{!}{
        \begin{tabular}{r|rrrr}
        \toprule
        \diagbox[width=12mm, height=6mm]{$T$}{$\lambda$} & 0.4   & 0.6   & 0.8   & 1 \\
        \midrule
        0.5   & 2.746 & 2.951 & 3.244 & 0.590 \\
        1     & 1.771 & 1.863 & 1.869 & \underline{0.521} \\
        3     & 0.628 & 0.623 & 0.597 & 0.737 \\
        5     & 0.533 & 0.599 & 0.720  & 0.901 \\
        \bottomrule
        \end{tabular}%
        }
        \label{tab:cifar10_ptb_normal_train}%
    \end{minipage} \quad \quad \quad 
    \centering
    \begin{minipage}[b]{.45\textwidth}
        \centering
        \caption{
            Average distance for normally trained model on CIFAR-10's non-member dataset.
        }
        \resizebox{.8\textwidth}{!}{
        \begin{tabular}{r|rrrr}
        \toprule
        \diagbox[width=12mm, height=6mm]{$T$}{$\lambda$} & 0.4   & 0.6   & 0.8   & 1 \\
        \midrule
        0.5   & 2.753 & 2.970  & 3.246 & 0.554 \\
        1     & 1.750  & 1.830  & 1.854 & \underline{0.509} \\
        3     & 0.612 & 0.606 & 0.586 & 0.732 \\
        5     & 0.533 & 0.600   & 0.722 & 0.904 \\
        \bottomrule
        \end{tabular}%
        }
        \label{tab:cifar10_ptb_normal_test}%
    \end{minipage}

    \centering
    \begin{minipage}[b]{.45\textwidth}
        \centering
        \caption{
            Average distance for normally trained model on CIFAR-100's member dataset.
        }
        \resizebox{.8\textwidth}{!}{
        \begin{tabular}{r|rrrr}
        \toprule
        \diagbox[width=12mm, height=6mm]{$T$}{$\lambda$} & 0.4   & 0.6   & 0.8   & 1 \\
        \midrule
        0.5   & 7.667 & 7.833 & 8.248 & 0.754 \\
        1     & 5.375 & 5.663 & 6.076 & \underline{0.628} \\
        3     & 2.557 & 2.464 & 1.607 & 1.373 \\
        5     & 1.292 & 1.045 & 1.442 & 2.151 \\
        \bottomrule
        \end{tabular}%
        }
        \label{tab:cifar100_ptb_normal_train}%
    \end{minipage} \quad \quad
    \centering
    \begin{minipage}[b]{.45\textwidth}
        \centering
        \caption{
            Average distance for normally trained model on CIFAR-100's non-member dataset.
        }
        \resizebox{.8\textwidth}{!}{
        \begin{tabular}{r|rrrr}
        \toprule
        \diagbox[width=12mm, height=6mm]{$T$}{$\lambda$} & 0.4   & 0.6   & 0.8   & 1 \\
        \midrule
        0.5   & 7.620  & 7.804 & 8.243 & 0.687 \\
        1     & 5.376 & 5.665 & 6.072 & \underline{0.628} \\
        3     & 2.715 & 2.719 & 1.770  & 1.355 \\
        5     & 1.263 & 1.047 & 1.373 & 2.112 \\
        \bottomrule
        \end{tabular}%
        }
        \label{tab:cifar100_ptb_normal_test}%
    \end{minipage}
\end{table*}%

\subsection{Adversarial Perturbation Distance Based Approach}
Clearly, the above strawman approach $I_{sm\_adv}$ has one serious drawback: it does not use the information about the size of the adversarial perturbation. This can be amended by explicitly defining a perturbation distance, and comparing the distance with a threshold like other metric-based methods discussed in Sec.~\ref{sec:mbmi}.

The perturbation distance between the example $x$ and the decision boundary can be approximated by
\begin{IEEEeqnarray}{C}
    \Dist(x) =  \| \hat{x}-x \|_2 \\ 
    \mathrm{~subject~to~} (\argmax F(x)) \neq (\argmax F(\hat{x})), \nonumber 
\end{IEEEeqnarray}
where $\hat{x}$ is the adversarial version associated with $x$. Obviously, the direct usage of $\Dist(x)$ as the metric does not guarantee the best membership inference performance, since $\Dist(x)$ can only approximate the distance between $x$ and the decision boundary. 

As an improvement, we set the distance between the example $x_i$ and the decision boundary along the direction specified by $y_t$ as
\begin{IEEEeqnarray}{C}
    \Dist(x_i, y_t) = \argmin_{ \| \hat{x}_i-x_i \|_2 } \Big( F(\hat{x}_i)=y_t \Big), 
\end{IEEEeqnarray}
where $F(x)=y\neq y_t$.
Under the white-box setting and use $y_t$'s the one-hot encoding label $p_n(j)$ ($t=\argmax p_n(j)$), the C\&W attack can be adopted to produce the optimal adversarial perturbation $\Dist(x_i, y_t)$ as follows:
\begin{IEEEeqnarray}{c}
    \argmin_{ \| \hat{x}_i-x_i \|_2 } \Big( \alpha \cdot \mathcal{L}_{CE}\big(F(\hat{x}_i), p_n(j)\big)  + \| \hat{x}_i-x_i \|_2 \Big).  
    \label{eq:dist_p}
\end{IEEEeqnarray}

Resort to the assumption that member examples are be more robust than non-members, the inference attack based on the perturbation distance $\Dist(x_i, y_t)$ can be defined as 
\begin{IEEEeqnarray}{rCl}
    I_{adv} (x_i) = \mathbb{I}\Big( \Dist(x_i, y_t)>\tau \Big).
    \label{Eq:infadv}
\end{IEEEeqnarray}
We remark $I_{adv}$ is the upper-bound benchmark for studying membership inference attacks, no matter the attack is white-box/black-box or is neural-based/metric-based \cite{Shokri2017MembershipIA,Yeom2018PrivacyRI,Nasr2019ComprehensivePA,Leino2020StolenML,Sablayrolles2019WhiteboxVB, Song2020SystematicEO,Li2020MembershipLL,ChoquetteChoo2020LabelOnlyMI}.


Recall the result shown in Fig.~\ref{Fig:direction}, adversarial direction indeed matters when inferring an example is a member or not. However, when only one-hot encoded label is used, it is not necessary that the perturbation distances of members and non-members are clearly separable along the direction specified by the one-hot encoded label. In view of this, we make use of the smoothed label $q_n(\lambda, j)$ to replace the one-hot encoded label $p_n(j)$, so the adversarial directions changes when varying the parameter $\lambda$. The new distance is now defined as 
\begin{IEEEeqnarray}{rCl}
&&\Dist(x_i, q_n(\lambda, j)) \nonumber \\
    =&& \argmin_{ \| \hat{x}_i-x_i \|_2 } \Big( \alpha \mathcal{L}_{CE}\big(F(\hat{x}_i), q_n(\lambda, j)\big)  \nonumber + \| \hat{x}_i-x_i \|_2  \Big)
\end{IEEEeqnarray}

\begin{table*}
    \centering
    \begin{minipage}[b]{.45\textwidth}
        \centering
        \caption{Inference accuracy for normally trained model on Fashion.}
        \resizebox{.8\textwidth}{!}{
        \begin{tabular}{r|rrrr}
        \toprule
        \diagbox[width=12mm, height=6mm]{$T$}{$\lambda$} & 0.4   & 0.6   & 0.8   & 1 \\
        \midrule
        0.5   & 0.528 & 0.509 & 0.500 & 0.526 \\
        1     & 0.518 & 0.509 & 0.514 & \underline{0.532} \\
        3     & 0.520 & 0.516 & 0.528 & 0.523 \\
        5     & 0.532 & 0.534 & \textbf{0.537} & 0.532 \\
        \bottomrule
        \end{tabular}%
        }
        \label{tab:fashion_mi_normal}%
    \end{minipage} \quad \quad \quad 
    \centering
    \begin{minipage}[b]{.45\textwidth}
        \centering
        \caption{
        Inference accuracy for normally trained model on CIFAR-10.}
        \resizebox{.8\textwidth}{!}{
        \begin{tabular}{r|rrrr}
        \toprule
        \diagbox[width=12mm, height=6mm]{$T$}{$\lambda$} & 0.4   & 0.6   & 0.8   & 1 \\
        \midrule
        0.5   & 0.511 & 0.507 & 0.515 & \textbf{0.576} \\
        1     & 0.526 & 0.520 & 0.528 & \underline{\textbf{0.576}} \\
        3     & 0.545 & 0.556 & 0.575 & 0.564 \\
        5     & 0.560 & 0.564 & 0.544 & 0.537 \\
        \bottomrule
        \end{tabular}%
        }
        \label{tab:cifar10_mi_normal}%
    \end{minipage} 

    \centering
    \begin{minipage}[b]{.45\textwidth}
        \centering
        \caption{Inference accuracy for normally trained model on CIFAR-100.}
        \resizebox{.8\textwidth}{!}{
        \begin{tabular}{r|rrrr}
        \toprule
        \diagbox[width=12mm, height=6mm]{$T$}{$\lambda$} & 0.4   & 0.6   & 0.8   & 1 \\
        \midrule
        0.5   & 0.526 & 0.536 & 0.519 & 0.557 \\
        1     & 0.506 & 0.506 & 0.522 & \underline{0.520} \\
        3     & 0.505 & 0.500 & 0.501 & 0.513 \\
        5     & \textbf{0.582} & 0.537 & 0.572 & 0.546 \\
        \bottomrule
        \end{tabular}%
        }
        \label{tab:cifar100_mi_normal}%
    \end{minipage}
\end{table*}%

At the same time, to match the directional information provided by the smoothed label $q_n(\lambda, j)$, we incorporate a temperature parameter $T$ into the softmax function of the target model $F$, which is defined as
\begin{IEEEeqnarray}{rCl}
    \softmax(x,T)_l = \frac{e^{x_l/T}}{\sum_j e^{x_j/T}},
\end{IEEEeqnarray}
for $l \in [1,n]$. From above notation, it is obvious that $\softmax(x,T) = \softmax(x/T)$. 
Therefore, the model uses the modified $\softmax$ becomes
\begin{IEEEeqnarray}{rCl}
    F(x, T) &=& \softmax(Z(x), T) \nonumber\\
            &=& \softmax(Z(x)/T).
\end{IEEEeqnarray}
\noindent \textbf{Remark.} 
The introduction of a temperature parameter, $T$, is motivated by the fact that, as demonstrated in \cite{Wang2021KnowledgeDA}, a temperature parameter aids in controlling the magnitude of the entries of the probability vector $F(x, T)$.
A larger value for $T$ produces a flattened probability distribution over all classes.  
Though being flattened, it is obvious that $(\argmax F(x, T)) = (\argmax F(x))$, implying the fact that the original classification functionality is well preserved. 


By incorporating the label smoothing technique and the temperature, the directional adversarial distance can be obtained as
\begin{IEEEeqnarray}{rCl}
    &&\Dist(x_i, q_n(\lambda, j), T) 
    \nonumber \\
    &=&\argmin_{ \| \hat{x}_i-x_i \|_2 } \Big( \alpha \mathcal{L}_{CE}\big(F(\hat{x}_i, T ), q_n(\lambda, j)\big)  
    \nonumber \\ 
    &+& \| \hat{x}_i-x_i \|_2 \Big). 
\end{IEEEeqnarray}
In particular, the state-of-the-art benchmark inference metric $\Dist(x_i, p_n(j))$ is just a special case of $\Dist(x_i, q_n(\lambda, j), T)$ for $\lambda=T=1$. 

Taking the directional distance $\Dist(x_i, q_n(\lambda, j), T)$ as a new metric, we propose an inference strategy $I_{dd}$, shown as follows
\begin{IEEEeqnarray}{L}
    I_{dd} (x_i, T, \lambda) = \mathbb{I}\Big( \min_{t}( \Dist(x_i, q_n(\lambda, j), T ) )>\tau \Big), 
    \IEEEeqnarraynumspace
    \label{Eq:iddinfer}
\end{IEEEeqnarray}
where $t=\argmax q_n(\lambda, j)$ is the target class of adversarial attack. 
In particular, $I_{adv}$ is a special case of $I_{dd}$ for $\lambda=T=1$.

\begin{table}
\centering
\begin{minipage}[b]{.45\textwidth}
    \centering
    \caption{
        Model accuracy after PGD adversarial training.}
    \resizebox{0.9\textwidth}{!}{
      \begin{tabular}{l|rrr}
      \toprule
            & \multicolumn{1}{c}{Fashion} & \multicolumn{1}{c}{CIFAR-10} & \multicolumn{1}{c}{CIFAR-100} \\
      \midrule
      Normal acc. & 0.861 & 0.981 & 0.695 \\
      Adv. acc. & 0.763 & 0.785 & 0.296 \\
      \bottomrule
      \end{tabular}%
    }
    \label{tab:pgd_performance}%
    \end{minipage}
\end{table}%

\begin{table*}
    \centering
    \begin{minipage}[b]{.45\textwidth}
        \centering
        \caption{
            Average distance for PGD-AT trained model on Fashion’s member dataset.
        }
        \resizebox{.8\textwidth}{!}{
            \begin{tabular}{r|rrrr}
                \toprule
                \diagbox[width=12mm, height=6mm]{$T$}{$\lambda$} & 0.4   & 0.6   & 0.8   & 1 \\
                \midrule
                0.5   & 4.255 & 4.434 & 4.413 & 3.961 \\
                1     & 3.672 & 4.003 & 4.205 & \underline{4.666} \\
                3     & 2.661 & 3.133 & 3.819 & 4.802 \\
                5     & 2.296 & 2.807 & 3.536 & 4.353 \\
                \bottomrule
            \end{tabular}%
        }
        \label{tab:fashion_ptb_pgd_train}
    \end{minipage} \quad \quad
    \centering
    \begin{minipage}[b]{.45\textwidth}
        \centering
        \caption{
            Average distance for PGD-AT trained model on Fashion’s non-member dataset.
        }
        \resizebox{.8\textwidth}{!}{
            \begin{tabular}{r|rrrr}
                \toprule
                \diagbox[width=12mm, height=6mm]{$T$}{$\lambda$} & 0.4   & 0.6   & 0.8   & 1 \\
                \midrule
                0.5   & 4.319 & 4.481 & 4.439 & 4.000 \\
                1     & 3.652 & 3.980 & 4.193 & \underline{4.647} \\
                3     & 2.659 & 3.127 & 3.812 & 4.790 \\
                5     & 2.289 & 2.802 & 3.529 & 4.357 \\
                \bottomrule
            \end{tabular}%
        }
        \label{tab:fashion_ptb_pgd_test}%
    \end{minipage}
    \centering
    \begin{minipage}[b]{.45\textwidth}
        \centering
        \caption{
            Average distance for PGD-AT trained model on CIFAR-10’s member dataset.
        }
        \resizebox{.8\textwidth}{!}{
        \begin{tabular}{r|rrrr}
        \toprule
        \diagbox[width=12mm, height=6mm]{$T$}{$\lambda$} & 0.4   & 0.6   & 0.8   & 1 \\
        \midrule
        0.5   & 2.226 & 2.227 & 2.190 & 1.847 \\
        1     & 1.903 & 1.890 & 1.855 & \underline{1.774} \\
        3     & 1.393 & 1.466 & 1.555 & 1.684 \\
        5     & 0.986 & 1.069 & 1.180 & 1.317 \\
        \bottomrule
        \end{tabular}%
        }
        \label{tab:cifar10_ptb_pgd_train}%
    \end{minipage} \quad \quad
    \centering
    \begin{minipage}[b]{.45\textwidth}
        \centering
        \caption{
            Average distance for PGD-AT trained model on CIFAR-10’s non-member dataset.
        }
        \resizebox{.8\textwidth}{!}{
        \begin{tabular}{r|rrrr}
        \toprule
        \diagbox[width=12mm, height=6mm]{$T$}{$\lambda$} & 0.4   & 0.6   & 0.8   & 1 \\
        \midrule
        0.5   & 2.007 & 1.997 & 1.951 & 1.558 \\
        1     & 1.729 & 1.715 & 1.678 & \underline{1.599} \\
        3     & 1.188 & 1.269 & 1.372 & 1.526 \\
        5     & 0.844 & 0.943 & 1.070 & 1.225 \\
        \bottomrule
        \end{tabular}%
        }
        \label{tab:cifar10_ptb_pgd_test}%
    \end{minipage}

    \centering
    \begin{minipage}[b]{.45\textwidth}
        \centering
        \caption{
            Average distance for PGD-AT trained model on CIFAR-100’s member dataset.
        }
        \resizebox{.8\textwidth}{!}{
        \begin{tabular}{r|rrrr}
        \toprule
        \diagbox[width=12mm, height=6mm]{$T$}{$\lambda$} & 0.4   & 0.6   & 0.8   & 1 \\
        \midrule
        0.5   & 3.459 & 3.387 & 3.220  & 2.766 \\
        1     & 3.000 & 3.054 & 3.133 & \underline{3.494} \\
        3     & 2.763 & 3.350 & 3.981 & 4.664 \\
        5     & 2.420 & 3.117 & 3.751 & 4.346 \\
        \bottomrule
        \end{tabular}%
        }
        \label{tab:cifar100_ptb_pgd_train}%
    \end{minipage} \quad \quad
    \centering
    \begin{minipage}[b]{.45\textwidth}
        \centering
        \caption{
            Average distance for PGD-AT trained model on CIFAR-100’s non-member dataset.
        }
        \resizebox{.8\textwidth}{!}{
        \begin{tabular}{r|rrrr}
        \toprule
        \diagbox[width=12mm, height=6mm]{$T$}{$\lambda$} & 0.4   & 0.6   & 0.8   & 1 \\
        \midrule
        0.5   & 3.464 & 3.376 & 3.211 & 2.762 \\
        1     & 2.862 & 2.890 & 2.967 & \underline{3.346} \\
        3     & 2.686 & 3.274 & 3.887 & 4.554 \\
        5     & 2.388 & 3.160 & 3.865 & 4.515 \\
        \bottomrule
        \end{tabular}%
        }
        \label{tab:cifar100_ptb_pgd_test}%
    \end{minipage}
\end{table*}%

\section{Experimental Results and Analyses}
\label{sec:exp}
\subsection{Experimental Settings}

We assess the performance of $I_{dd}$ on three datasets: Fashion-MNIST \cite{Xiao2017FashionMNISTAN}, CIFAR-10 \cite{Krizhevsky2009LearningML} and CIFAR-100 \cite{Krizhevsky2009LearningML}.  
The VGG-16 \cite{Simonyan2015VeryDC} is used for CIFAR-10 and CIFAR-100 and a 5-layer CNN is used for Fashion-MNIST.
For all three datasets in the experiments, the choice of $\lambda$ is from $[0.4,0.6,0.8,1]$, and choice of temperature $T$ is from $[0.3,1,3,5]$.
For Fashion-MNIST, we set $\alpha=0.01$; for CIFAR-10, $\alpha=0.001$; and for CIFAR-100, $\alpha=0.01$.
For each dataset, $1000$ random-chosen examples from training and test sets are used as members and non-members respectively, which serves as the ground-truth for measuring inference accuracy. 
For Fashion-MNIST and CIFAR-10, we measure the directional adversarial distance of a given example to all the other classes except itself, i.e., the number of choices for $t$ of Eq.~(\ref{Eq:iddinfer}) is $9$. For CIFAR-100, we randomly choose $10$ classes for measuring the directional adversarial distance, i.e., the number of choices for $t$ of Eq.~(\ref{Eq:iddinfer}) is $10$. Similar to all other metric-based membership inference {attacks \cite{Nasr2019ComprehensivePA,Leino2020StolenML,Sablayrolles2019WhiteboxVB, Song2020SystematicEO}}, we do not pay special attention to the choice of the threshold value $\tau$. In what follows, only the $\tau$ that leads to the best results is presented without explicit explanation.

\subsection{Experimental Result Analyses}

We start our analyses by reporting that the averaged adversarial distance indeed varies when changing the adversarial direction. For this aim, we define the averaged adversarial distance as 
\begin{equation}
    \Dist(\lambda, T)=\frac{1}{m \cdot |\{t\}|} \sum_{i=1}^{m} \sum\nolimits_{t} \Dist(x_i, q_n(\lambda, j), T),
    \label{eq:dist_lambda_t}
\end{equation}
where $t=\argmax q_n(\lambda, j)$, $|\{t\}| = 9$ for Fashion-MNIST/CIFAR-10 and $|\{t\}| = 10$ for CIFAR-100, $x_i$ is an example from either the ground-truth member/non-member sets, and $m=1000$ as we randomly choose $1000$ member and non-member examples. 

The results of the averaged adversarial distance are tabulated by  Tables~\ref{tab:fashion_ptb_normal_train}-\ref{tab:cifar100_ptb_normal_test} for different settings of $T$ and $\lambda$ over different datasets.  
The cases for $T = \lambda = 1$ (i.e., $I_{adv}$) is emphasized by underlined text. 
The general trend that can be observed from these tables is when reducing the value of $T$ and increasing the value of $\lambda$, averaged adversarial distance for both member and non-member becomes smaller. Moreover, for the case $T = \lambda = 1$, the difference between averaged adversarial distances associated with members and non-member are not always large, for example, in Fashion-MNIST and CIFAR-100 the difference is $0$. That said, the method $I_{adv}$ characterized by Eq.~(\ref{Eq:infadv}), which is the benchmark used for other metric-based inference attacks, performs bad over Fashion-MNIST and CIFAR-100. By choosing appropriate values of $\lambda$ and $T$, it is possible to outperform the benchmarking method $I_{adv}$.

We then validate the claim above by quantitatively studying the accuracy of $I_{dd}$ (and $I_{adv}$ when $T = \lambda = 1$). Following the literature studies \cite{Song2020SystematicEO, Nasr2019ComprehensivePA, Shokri2017MembershipIA}, we set the inference accuracy as 
\begin{IEEEeqnarray}{rCl}
    Acc_{Inf} & = & \frac{1}{2} \Big( \frac{\sum_{x \in D_{train}} I(x) }{ | D_{train} |  } + 1  - \frac{ \sum_{ x \in D_{test} } I(x) }{ | D_{test} |  } \Big),\nonumber \\ 
    \label{eq:mi_acc}
\end{IEEEeqnarray}
where $| D_{test} |$ and $| D_{train} |$ represent the size of test and training datasets respectively. 

The accuracy of $I_{dd}$ are listed in Tables~\ref{tab:fashion_mi_normal}- \ref{tab:cifar100_mi_normal} with different settings of $T$ and $\lambda$. 
It is clear from these tables the claim above is true: $I_{adv}$ does not guarantee best inference accuracy\footnote{We did not list the accuracy of other metric-based methods here as $I_{adv}$ is already the previously-known best.}.
It can be observed that the accuracy of $I_{dd}$ according to different settings of $T$ and $\lambda$. And there is no such set of universal values for these parameters ($T$ and $\lambda$) that can be effective for all datasets, each dataset has its own suitable parameters for best inference. 

\begin{table*}
    \centering
    \begin{minipage}[b]{.45\textwidth}
        \centering
        \caption{
        Inference accuracy for PGD-AT trained model on Fashion.}
        \resizebox{0.8\textwidth}{!}{
            \begin{tabular}{r|rrrr}
                \toprule
                \diagbox[width=12mm, height=6mm]{$T$}{$\lambda$} & 0.4   & 0.6   & 0.8   & 1 \\
                \midrule
                0.5   & 0.507 & 0.504 & 0.508 & 0.506 \\
                1     & \textbf{0.523} & 0.514 & 0.512 & \underline{0.509} \\
                3     & 0.520 & 0.516 & 0.522 & 0.517 \\
                5     & 0.514 & 0.503 & 0.506 & 0.507 \\
                \bottomrule
            \end{tabular}%
        }
        \label{tab:fashion_mi_pgd}%
    \end{minipage} \quad \quad
    \centering
    \begin{minipage}[b]{.45\textwidth}
        \centering
        \caption{
        Inference accuracy for PGD-AT trained model on CIFAR-10.}
        \resizebox{0.8\textwidth}{!}{
        \begin{tabular}{r|rrrr}
        \toprule
        \diagbox[width=12mm, height=6mm]{$T$}{$\lambda$} & 0.4   & 0.6   & 0.8   & 1 \\
        \midrule
        0.5   & 0.602 & 0.606 & 0.631 & \textbf{0.722} \\
        1     & 0.601 & 0.617 & 0.655 & \underline{0.694} \\
        3     & 0.670 & 0.700 & 0.712 & 0.709 \\
        5     & 0.692 & 0.708 & 0.708 & 0.696 \\
        \bottomrule
        \end{tabular}%
        }
        \label{tab:cifar10_mi_pgd}%
    \end{minipage}

    \centering
    \begin{minipage}[b]{.45\textwidth}
        \centering
        \caption{
        Inference accuracy for PGD-AT trained model on CIFAR-100.}
        \resizebox{0.8\textwidth}{!}{
        \begin{tabular}{r|rrrr}
        \toprule
        \diagbox[width=12mm, height=6mm]{$T$}{$\lambda$} & 0.4   & 0.6   & 0.8   & 1 \\
        \midrule
        0.5   & 0.522 & 0.518 & 0.530 & 0.556 \\
        1     & 0.558 & 0.557 & 0.546 & \underline{0.518} \\
        3     & 0.600 & 0.611 & 0.638 & \textbf{0.665} \\
        5     & 0.517 & 0.511 & 0.504 & 0.506 \\
        \bottomrule
        \end{tabular}%
        }
        \label{tab:cifar100_mi_pgd}%
    \end{minipage}
\end{table*}%

\subsection{Membership Inference for Adversarially Trained Model}  
\label{sec:mi_adv_result}
By appropriately using the directional information, we have shown that the newly proposed inference method $I_{dd}$ outperforms the known benchmark method $I_{adv}$. Both these two method are based on adversarial robustness, so it is interesting to see the answers of the following two questions: 
\begin{itemize}
    \item Will inference methods based on adversarial robustness still be useful if the neural model is adversarially trained at the first place?
    \item Will other metric-based inference methods outperform adversarial robustness-based methods if the neural model is adversarially trained?
\end{itemize}

To probe the answers of these two questions, we perform PGD-AT adversarial training discussed in Sec.~\ref{subsec:aeat} for Fashion-MNIST, CIFAR-10, and CIFAR-100, respectively, and the model accuracy are reported in Table~\ref{tab:pgd_performance}. Since PGD-AT training is harder than normal training, it is clear from this table that the model accuracy is sacrificed to various degree. However, according to \cite{Madry2018TowardsDL}, adversarial training improves model robustness with regard to adversarial attack. 

Similarly, Tables~\ref{tab:fashion_ptb_pgd_train}-\ref{tab:cifar100_ptb_pgd_test} list the the averaged adversarial distance defined by Eq.~(\ref{eq:dist_lambda_t}) on adversarial trained models for all the considered datasets.
In contrast to Tables~\ref{tab:fashion_ptb_normal_train}-\ref{tab:cifar100_ptb_normal_test}, it is clear that all the averaged distances for adversarial trained models are larger than those for normally trained models. This agrees with the fact that adversarial training improves model robustness. 
However, though the model robustness is improved, the specific averaged distances for member and non-member under a specific attack setting (i.e., different choices of $\lambda$ and $T$) are still different. In particular, the trend that smaller $T$ and larger $\lambda$ yields smaller averaged distance is the same as the one revealed by Tables~\ref{tab:fashion_ptb_normal_train}-\ref{tab:cifar100_ptb_normal_test}.
In view of these, we speculate that adversarial robustness based inference attacks is still as effective, if not more effective, over adversarially trained models.

We validate this speculation by quantitatively studying the inference accuracy defined by Eq.~(\ref{eq:mi_acc}). The Tables~\ref{tab:fashion_mi_pgd}-\ref{tab:cifar100_mi_pgd} list the inference accuracy of $I_{dd}$ for different settings of $\lambda$ and $T$. Compared to the results shown in Tables~\ref{tab:fashion_mi_normal}-\ref{tab:cifar100_mi_normal} about normally trained models, it is surprise to see that adversarially trained models, though they owns better robustness with regard to adversarial attacks, is more fragile to membership inference attacks. Moreover, it is also clear that 
by changing $\lambda$ and $T$ from $1$ to other conditions, inference accuracy $Acc_{inf}$ increases, which means $I_{adv}$ is still inferior to $I_{dd}$. But there is no universal settings of $\lambda$ and $T$, which produce best membership inference results, that work for different datasets. 

Last but not least, we compare the performance of $I_{dd}$ with other metric-based membership inference methods over adversarially trained models, and the results are tabulated in Table~\ref{tab:mi_acc_pgd}. 
Clearly, $I_{dd}$ outperforms all other methods in all datasets. To summarize and answer the two questions mentioned at the beginning of the section: 
\begin{itemize}
    \item Adversarial robustness based inference methods are still useful even if the neural model is adversarially trained in the first place. And indeed, the adversarially trained model is more fragile to membership inference attacks. 
    \item Even if the model is adversarially trained, the performance of adversarial robustness-based inference methods is generally better than other metric-based inference methods because they explicitly employ the decision boundary information, which relates back to the fundamental rationale of membership inference: model overfitting \cite{Yeom2018PrivacyRI}. Among all known adversarial robustness based inference methods, the proposed $I_{dd}$ performs the best. 
\end{itemize}

\begin{table}
\centering
\begin{minipage}[b]{.45\textwidth}
    \centering
    \caption{Inference accuracy of different methods over adversarial trained models.}
    \resizebox{.9\textwidth}{!}{
    \begin{tabular}{l|rrr}
        \toprule
              & \multicolumn{1}{c}{Fashion} & \multicolumn{1}{c}{CIFAR-10} & \multicolumn{1}{c}{CIFAR-100} \\
        \midrule
        $I_{dd}$  (ours) & \textbf{0.523} & \textbf{0.722} & \textbf{0.665} \\
        $I_{sm\_adv}$ & 0.508 & 0.714 & 0.563 \\
        $I_{bl}$ & 0.509 & 0.596 & 0.595 \\
        $I_{cfs}$ & 0.514 & 0.564 & 0.610 \\
        $I_{CE}$ & 0.514 & 0.564 & 0.610 \\
        $I_{entropy}$ & 0.520  & 0.578 & 0.578 \\
        $I_{m\_entropy}$ & 0.513 & 0.568 & 0.607 \\
        \bottomrule
        \end{tabular}%
        }
    \label{tab:mi_acc_pgd}%
    \end{minipage}
\end{table}%

\section{Conclusion}
\label{sec:conclusion}

This paper presents a new technique $I_{dd}$ to improve the existing adversarial robustness-based membership inference method by adjusting the directions of adversarial perturbations. 
By evaluating the performance of normally trained and adversarially trained models under white-box settings and comparing $I_{dd}$ with other metric-based methods, we conclude that $I_{dd}$ can improve the performance of adversarial robustness based membership inference, and $I_{dd}$ outperforms all existing metric-based methods for adversarial robust models. 
This makes $I_{dd}$ a supplement to existing attack techniques to study the privacy leakage of deep learning.

\section*{Acknowledge}
This work was supported by the Technology Innovation and Application Development of Chongqing Science and Technology Commission (cstc2019jscx-kjfp0004, cstc2020jscx-msxm0917).

\section*{Data Availability}
The data underlying this article are available at \url{https://www.cs.toronto.edu/~kriz/cifar.html} and \url{https://github.com/zalandoresearch/fashion-mnist}.

\bibliographystyle{compj}
\bibliography{ref}

\end{document}